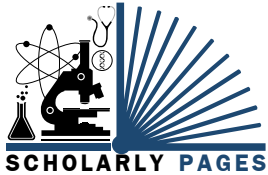

# Trends in Artificial Intelligence

**Research Article**  **Open Access**

# Scan Transcription of Two-Dimensional Shapes as an Alternative Neuromorphic Concept

*Ernest Greene*[*] *and Yash Patel*

*Department of Psychology, Neurometric Research Laboratory, University of Southern California, Los Angeles, USA*

*A similarity response is only one of many kinds of response we can elicit from a subject, by suitable manipulation of the number and variety of the stimuli presented, and by instructions to make one sort of decision or choice rather than another.* Robert A Gregson [1]

Selfridge [2,3], along with Sutherland [4] and Marr [5,6] provided some of the earliest proposals for how to program computers to recognize shapes. Their emphasis on filtering for contour features, especially the orientation of boundary segments, was reinforced by the Nobel Prize winning work of Hubel & Wiesel [7,8] who discovered that neurons in primary visual cortex selectively respond as a function of contour orientation. Countless investigators and theorists have continued to build on this approach, with references [9-16] providing a small sampling of recent work. These models are often described as "neuromorphic", which implies that the computational methods are based on biologically plausible principles. Recent work from the present lab has challenged the emphasis on orientation selectivity and the use of neural network principles [17]. The goal of the present report is not to re-litigate those issues, but to provide an alternative concept for encoding of shape information that may be useful to neuromorphic modelers.

Several lines of evidence have converged to suggest that basic shape encoding is based on eye motion or has at least is derived from motion-related mechanisms. Ahissar & Arieli [18] proposed that during fixation the eyes drift cyclically for a few hundred milliseconds, which converts spatial signals into a temporal code. Rucci & Victor [19] stress that fixational drift enhances high spatial frequencies and thus improves visibility. Gollisch & Meister [20] suggest that contours trigger synchronous firing, i.e. a latency code, from retinal ganglion cells at the termination of saccades.

The concept that latency of firing could encode stimulus information was advanced by Bullock [21] and elaborated by Hopfield [22] among others [23-26]. In theory, the successive waves of elicited spikes can travel up through successive neuronal populations quickly, providing for very fast discrimination or recognition. There is evidence for rapid discrimination and recognition that would require a fast and efficient mechanism. For example, Thorpe and associates [23] displayed an inventory consisting of thousands of slides. Each slide either included the image of an animal or did not, and observers could accurately report that an animal was present with 94% accuracy. Further, measures of event-related potentials recorded from frontal areas of the brain found a signal that predicted which choice the observer would make, and the decision for whether the animal was present was elicited within 150 ms of the display time. As detailed by the authors, the stimulus cues would need to pass through a number of neuronal populations in that time, which argues against a rate code and in favor of a latency (population) code. Similar results requiring recognition of other kinds of objects were subsequently reported [24-26].

While a latency (population) code might well contribute to the speed of signal transmission, here we are emphasizing a different potential benefit, which is to provide a shape-encoding method that does not require extensive training trials. A process described as "scan transcription" mimics the action of stimulus or eye motion, eliciting spikes from retinal ganglion cells as the wave passes across the shape contours. Locations on the outer boundary are especially critical shape features, and it is convenient to describe these as "boundary markers". As the wave passes across these boundary markers, it generates a differential density of spikes as a function of the number of markers that were encountered at successive moments in time. This information can be used to summarize, store, and allow for subsequent identification of a given shape.

In the sections that follow, we first simulate the scan-transcription process, and convert this information into a summary histogram. Summary histograms were derived for an inventory of unknown shapes, each of which was a continuous string of dots forming an outer boundary and designed to not look like known objects. Paired comparison of these summaries provided shape-difference scores. Then we evaluated whether the shape-difference scores could predict judgments that were provided in a shape-matching task.

### Simulation of the Scan-Transcription Process

Here we simulated the concept that motion or motion-derived mechanisms, i.e. scans, can provide shape information that allows for identification and/or assessment of shape similarity. Each shape in a 480-shape inventory was represented as a string of discrete locations on a 64 × 64 array of dots, this corresponding to the displays that would provide data on match recognition, described more fully below. Figure 1 illustrates an example of a shape being scanned to register the encounter of boundary markers. Various scan sequences might be used, but here we counted the number of dots encountered from successive scans of all columns and rows. The column and row counts were placed in tandem to provide 128-bin histograms. These were then trimmed to eliminate any empty bins, i.e. these being the columns or rows of the background for each shape. Then the values are re-binning into 20 bins and normalized, thus yielding a summary for each shape that could be used for identification or similarity comparison.

### Computational Scaling of Shape Similarity

Members of the shape inventory were paired, and the (480 choose 2) combinations provided for 114,960

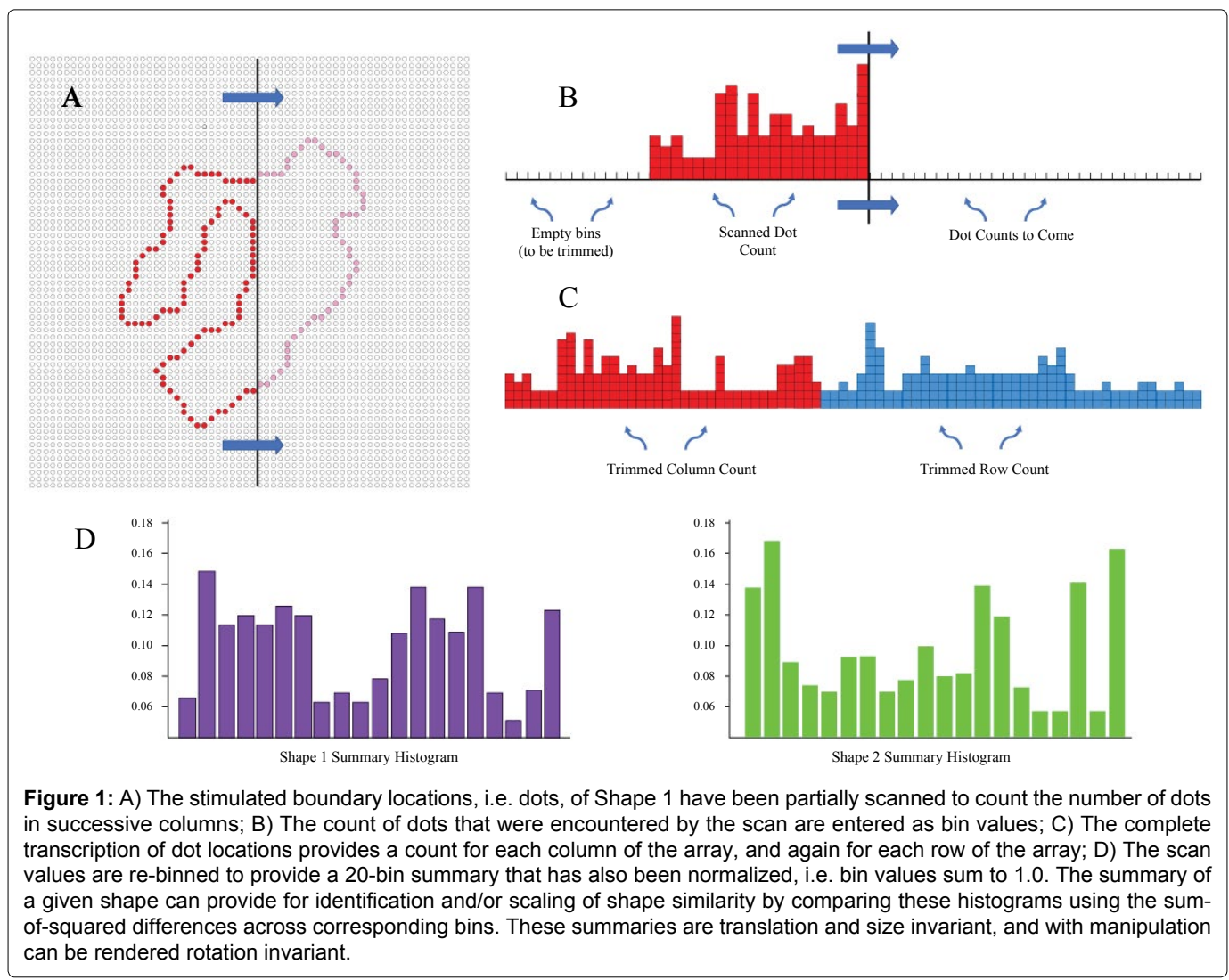

**Figure 1:** A) The stimulated boundary locations, i.e. dots, of Shape 1 have been partially scanned to count the number of dots in successive columns; B) The count of dots that were encountered by the scan are entered as bin values; C) The complete transcription of dot locations provides a count for each column of the array, and again for each row of the array; D) The scan values are re-binned to provide a 20-bin summary that has also been normalized, i.e. bin values sum to 1.0. The summary of a given shape can provide for identification and/or scaling of shape similarity by comparing these histograms using the sum-of-squared differences across corresponding bins. These summaries are translation and size invariant, and with manipulation can be rendered rotation invariant.

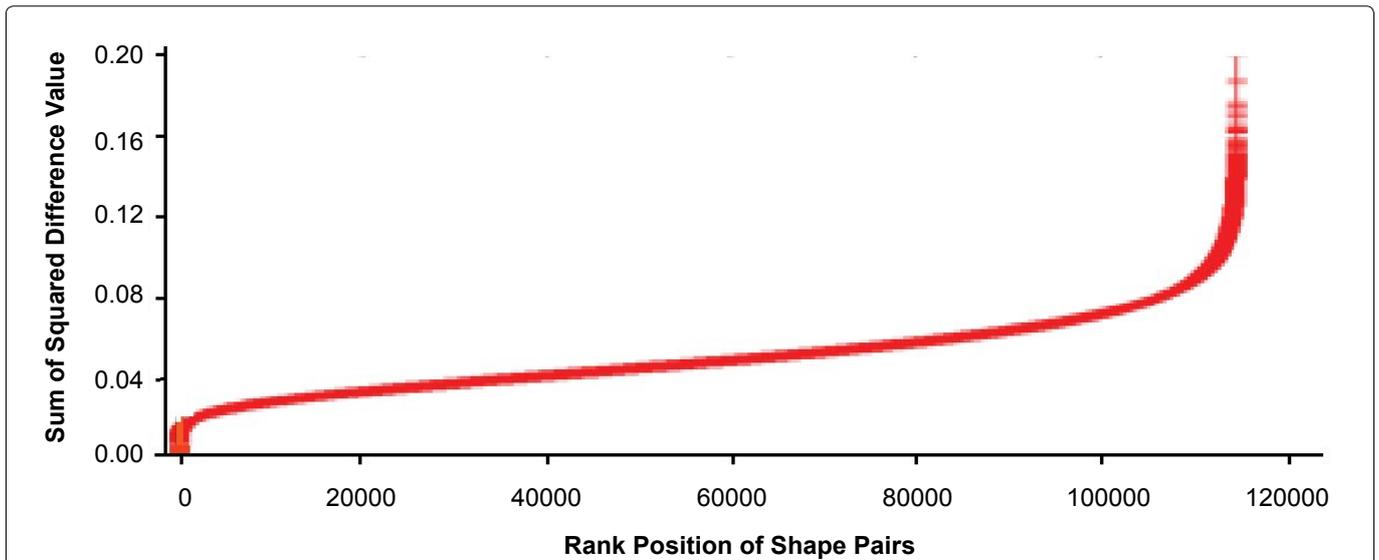

**Figure 2:** Ranked sum-of-squared difference values. The ranking of SSD values provided linear increments across a large range of the pairings. These values can be designated as "shape-difference" scores.

pairs. The histograms for each pair were then compared using a Sum-of-Squared-Differences calculation (SSD). The resulting SSD values could be used for shape-identification, in that no pairing yielded an SSD of zero except when the histograms were both from the same shape.

However, the goal here was not to prove the method's value as a recognition tool but to evaluate shape similarity. Towards that end, the SSD values were ranked for size, thus providing a similarity scale as shown in Figure 2.

The expectation was that SSD values would be smallest for pairs in which the members were similar, and would be larger for those that were less similar. To keep the size terminology from being confusing, it is best to describe the value as reflecting the "difference" in similarity.

### Evaluating Similarity Scaling with a Shape-Matching Task

The SSD calculation provides quantitative evidence that the summary histograms are similar, but this does not prove that scan transcription is capturing what humans would see as similar shape features. To provide that evidence, a match recognition experiment was conducted to determine whether the size of the shape-difference scores would predict the frequency with which respondents judge pair members as being the "same".

The present experiment follows on work from this laboratory that asked respondents to identify unknown shapes that were seen only once [17]. The reported experiments briefly flashed the boundary dots on an LED display board, first showing a target shape, followed shortly after by a comparison shape that was either a low-density (sparse-dot) version of the target, or a low-density version of a different shape. Using a low-density version of the comparison shape was needed to preclude ceiling effects, i.e. respondents making few if any judgment errors. Respondents could identify which trials provided a low-density version of the target at a level that was well above chance even when the density of boundary dots was decreased to 5%, and also manifested translation-, size-, and rotation-invariance in the judgments [17].

The present experiment used the same display system and test protocols as in the earlier report - see [17] for details. The major differences were in using mostly non-matching pairs as the target and comparison shapes, and assessing whether the size of the shape-difference value would differentially predict the probability that pairs would be judged as being the "same".

Three-hundred pairs at equal intervals along the scale were chosen for testing in the matching task. To assure that the relative abundance of dots would not be a factor in the judgment, the number of dots forming the boundary of pairs did not differ by more than 10. Ninety additional shapes were chosen for trials wherein the comparison shape was a low-density version of the target shape - these being described as "same-shape pairs".

Each of the eight respondents judged the 390 pairs in a single session. Each respondent was informed that each trial would display two shapes, the first being considered as the "target" and the second was a "comparison shape". They were specifically informed that the comparison shape might have been derived from the target, with some of the boundary dots deleted, or might be a sparse-dot version of a different shape. Their task was to say whether the two were the "same" or "different". For convenience, we are describing this as a "similarity judgment". Each was assured that if the comparison shape was derived from the target, it would not be altered in its orientation, i.e. the shape would not be rotated.

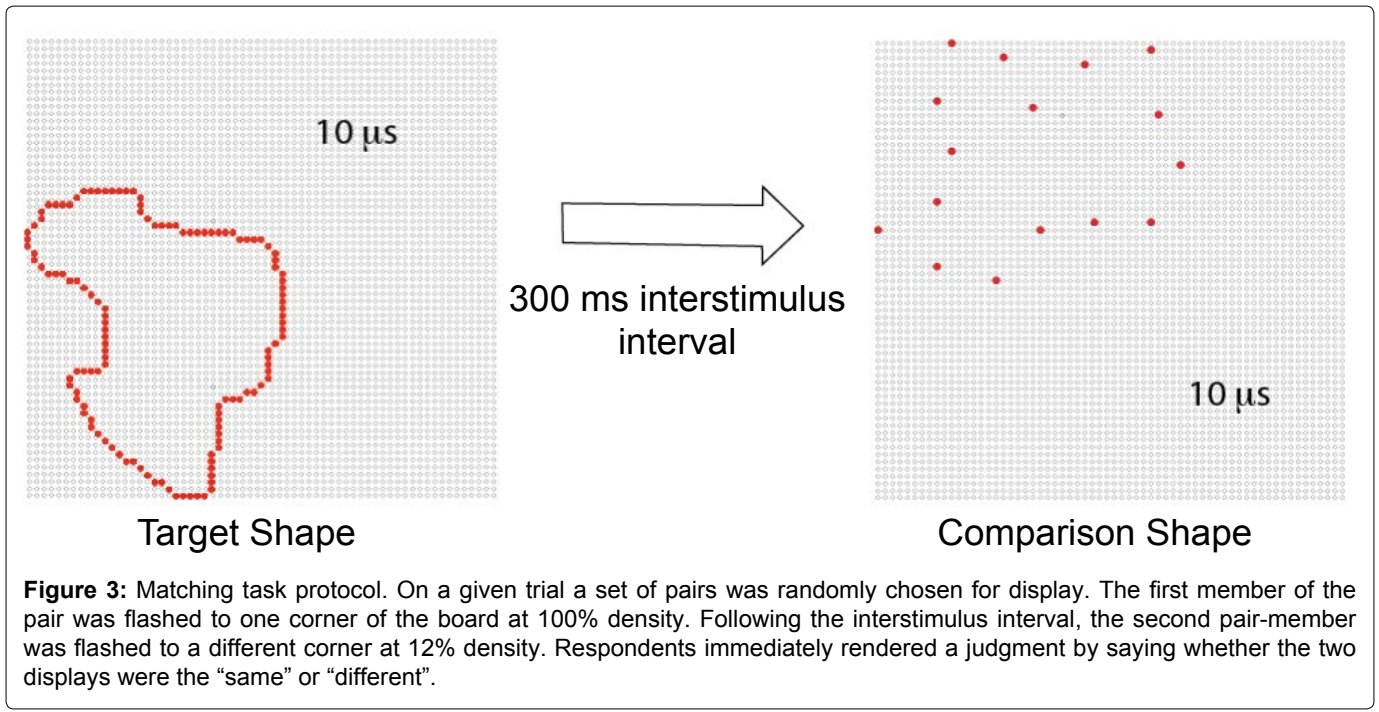

**Figure 3:** Matching task protocol. On a given trial a set of pairs was randomly chosen for display. The first member of the pair was flashed to one corner of the board at 100% density. Following the interstimulus interval, the second pair-member was flashed to a different corner at 12% density. Respondents immediately rendered a judgment by saying whether the two displays were the "same" or "different".

Pairs were drawn from the display set at random, with one member of the pair being displayed at 100% density and the other member being displayed at 12% density. To provide this density, on each trial a random dot was picked and then every 8th dot in the boundary sequence was displayed, skipping an additional dot as needed to maintain the specified percentage. On each trial a fixation dot was provided for the duration of 500 ms, after which the target shape, i.e. the first pair member, was flashed for 10 μs. The target was displayed in a corner of the board that was randomly chosen, such that one or more boundary dots were located on an outer column and row of the dot array. After a 300 ms interstimulus interval the comparison shape, i.e. the second pair member, was flashed for 10 μs in a different corner of the board, chosen at random. The sequence is illustrated in Figure 3.

After the pair was displayed, the respondent immediately said either "same" or "different" to indicate whether the comparison shape was judged to be derived from the target shape. The response was recorded by the experimenter using an on-screen button, that action serving to launch the next trial. Neither the experimenter nor the respondent was provided with information as to which pair had been displayed, nor was any feedback on the response provided.

## Shape Difference Values Predict Matching Judgments

The 300 pair judgments were analyzed with mixed-effect logistic regression, binary choices versus the difference value, which provided models for each respondent as well as group models. The means of judgments for the 90 same-shape pairs were available as anchors, i.e. indicating the maximum level that might be expected for pairs that had low shape difference values. These means were not included in the regressions.

The regression model for each respondent had a significant linear term ($p = 0.0002$ for one respondent and $p < 0.0001$ for the other seven). Quadratic and cubic terms were not significant and have not been included in the plots shown in Figure 4A. Note that each model curves a bit even though only the linear term was used, which is a characteristic of logistic linear regression. Means for same-shape judgments are shown at a shape difference value of zero.

The group model and confidence band is plotted in Figure 4B. This model had significant linear ($p < 0.0001$), quadratic ($p < 0.006$) and cubic terms ($p < 0.001$). This indicates that the small curvature differential seen in individual models were sufficiently consistent to yield significant quadratic and cubic components for the group model. Nonetheless, it is clear from the group plot (Figure 4B) that the linear component provides the major source of change in the similarity judgment as a function of shape difference value.

On average, respondents judged the pairs having the lowest shape difference values as being the same on just over 70% of the trials. This was not far below the level of performance for trials that displayed a low-density version of the target shape, i.e. what we have called the "same-shape judgments". The mean proportions of same-shape judgments for individual subjects and for the group are shown in the upper-left corner of each plot.

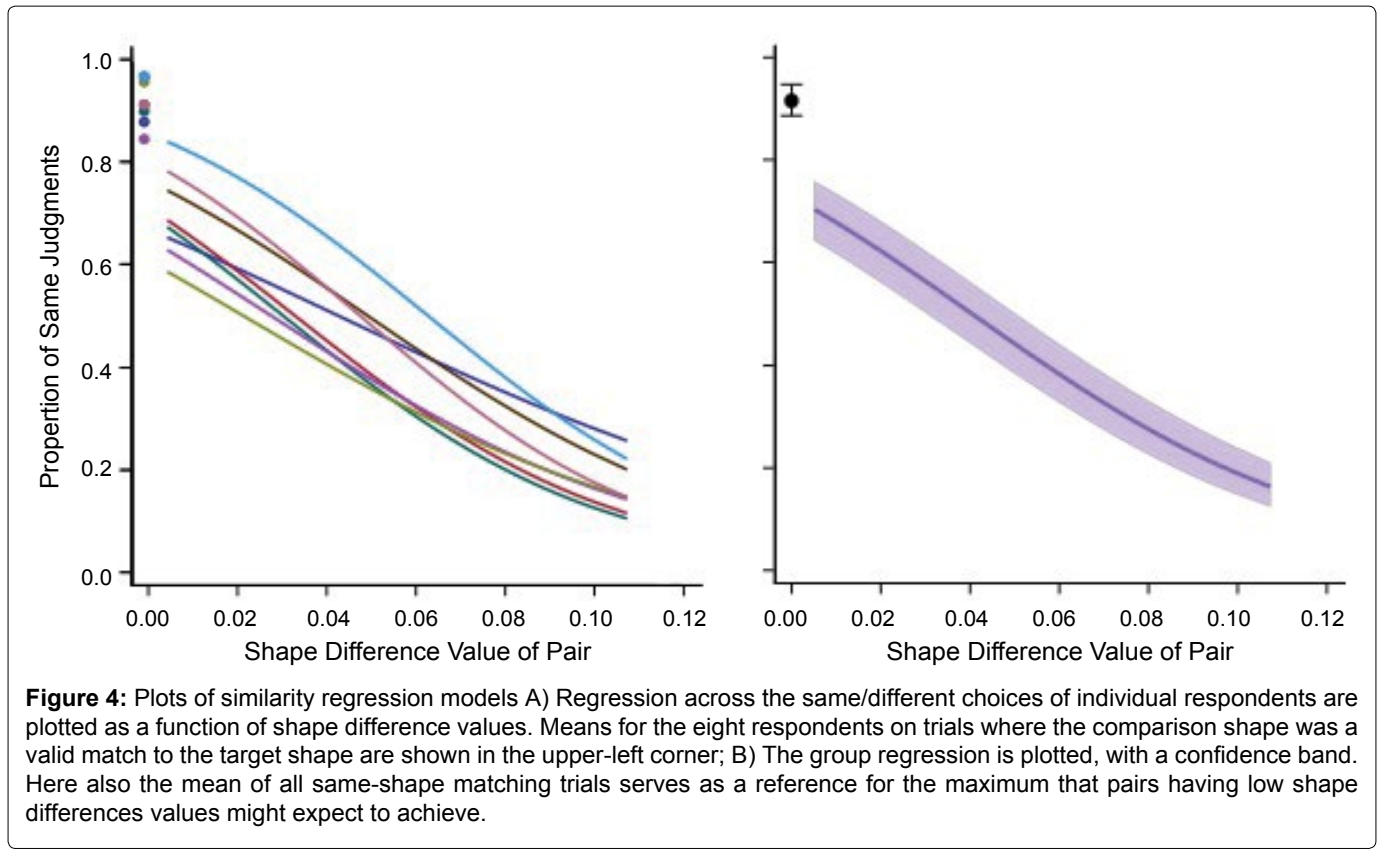

**Figure 4:** Plots of similarity regression models A) Regression across the same/different choices of individual respondents are plotted as a function of shape difference values. Means for the eight respondents on trials where the comparison shape was a valid match to the target shape are shown in the upper-left corner; B) The group regression is plotted, with a confidence band. Here also the mean of all same-shape matching trials serves as a reference for the maximum that pairs having low shape differences values might expect to achieve.

As shape difference values increased, there were progressively fewer decisions that the shapes appeared to be similar. The decline was monotonic across all eight subjects, and almost linear. For the largest shape difference values, the "same" judgments were offered on less than 20% of the trials.

These findings indicate that the scan protocol for summarizing shapes and specifying shape similarity is effective at predicting decisions in the matching judgment task. It should be emphasized that for 300 out of 390 trials the comparison shape was different from the target shape. Yet the respondents described the comparison shape as being the "same" on a great many of these trials, and the likelihood of that decision was heavily determined by the size of the shape difference value. This was the case even though respondents saw a given unknown shape only once (Figure 4).

It would be useful to re-affirm that respondents were told that the comparison shape could be a sparse version of the target shape, so that a "same" judgment meant that they thought it had been derived from the target shape by deleting some of the dots. In fact, a great majority of the comparison shapes - 77% - were from a different shape. These stimuli elicited a "same" judgment on almost half of the trials wherein the pairs were different. That, combined with the significant regression of matching judgments against shape difference values strongly implies that scan encoding provides a useful method for scaling shape similarity.

## Biological Plausibility of Scan Encoding

There are several excellent reviews of motion encoding systems [27-30], and others have discussed how motion could provide for activation of signals that register the contour features [18-20]. It serves no useful purpose to review those concepts again. Of greater relevance is the question of how the system could provide for latency encoding of boundary markers that are simultaneously displayed. If all the markers are activated at the same moment, as would be expected with a 10-microsecond flash, how could the alignments and relative number of adjacent markers can be converted into a latency code?

Our hypothesis is that primitive shape-encoding mechanisms based on stimulus motion or eye motion has evolved to provide for scanning of still images. This might be accomplished within the retina. Greene [31,32] discussed how polyaxonal amacrine (PA1) cells generate spreading waves that might serve to register distances among boundary-marked locations. The physiology of these neurons was first characterized in primate retina [33]. These cells have a relatively narrow dendritic field, which would provide for good resolution of contour markers. But unlike most other amacrine neurons, they have a widespread branching pattern of axons that extend across large areas [33-39]. In Macaque, the span of the axon field is roughly ten times larger than the dendritic field [33]. Converted to human units, the axonal arbor of a single PA1 neuron would be about 7.5 degrees of visual angle. Wright & Vaney [40] report that they appear to connect to a single class of retinal ganglion cell, these being "local edge detectors". The axons of a given PAI neuron are relatively sparse, but there is overlap of upward of 200-300 neurons [40], so the overall mesh of the axonal plexus can almost be described as the woof and weave of a fabric [32,39].

It has been suggested that the function of PA1 neurons is to suppress responding of retinal ganglion cells during rapid shifts of images, which may be specific to execution of saccadic eye movements [41,42]. But we are recommending a more specific role in controlling when stimulus content is delivered as part of a population code. Ackert and associates [43] reported that these cells couple with ON direction-sensitive ganglion cells via gap junctions to synchronize spontaneous and light-evoked spike activity of the neighboring cells in this class. This response provided synchronization for movement in all except the null direction, so it is unclear whether this would provide for selectivity of shape encoding. Kenyon and associates [44] included PA1 neurons in a model that provided control of synchronous firing for improvement of stimulus discrimination. This research group further evaluated how these neurons could provide for enhancing the encoding of large image features through high frequency resonance, as might be generated from eye tremor [45]. It should be clear that any control of oscillation must be accomplished through spreading waves, which allows for the possibility of polling that would provide for shape encoding.

## Neuromorphic Implications

Biological systems for processing image information are currently far more effective than the best systems that have been developed for computer vision. Given that, it is reasonable for engineers to ask for insights that might be offered by the neuroscience community. It is possible, however, that they have been ill served by the advice this community has provided. A great many models for how to encode shape information have been offered, most being connectionist (neural net) methods by which connections among several neuron populations are modified through many hundreds, sometimes thousands, of training trials [46-53]. Those many training trials are needed to adjust the balance of activation that provides for identification of alternative shapes as well as translation, rotation, and size invariance. The prior match-recognition experiments [17] demonstrated that humans can encode shapes that have been seen only once and can use that information within moments to identify whether the shape was seen previously. This argues that biological vision is not using connectionist methods to encode shapes, and new methods that can quickly summarize shapes are needed. Above we have argued that spreading retinal waves transcribe the locations of boundary markers into a temporal message, wherein the incidence of boundary encounter is communicated by means of a latency (population) response.

It should be clear, however, that there are numerous alternative means to derive this information. While the histogram counts illustrated in Figure 1 can serve as proxies for the number of simultaneous spikes from retinal ganglion cells, they could just as well be binary values delivered from the shift registers of a silicon retina. The discussion of retinal physiology should serve only as support for the biological plausibility of the concepts. It is assumed that engineers will have the best insights about how those concepts might be implemented for machine vision.

## Acknowledgments

The present research was supported by funding from the Quest for Truth Foundation.

## References

1. Gregson RA (1975) Psychophysics of similarity. Academic Press, New York, 15.

2. Selfridge OG (1957) Pattern recognition and learning. In: Cherry C, Information theory. Academic Press, New York, 345-353.

3. Selfridge OG (1959) Pandemonium: A paradigm for learning in the mechanization of thought process. HM Stationary Office, London.

4. Sutherland NS (1968) Outlines of a theory of visual pattern recognition in animals and man. Proc R Soc Lond B Biol Sci 171: 297-317.

5. Marr D, Nishihara HK (1978) Visual information processing, artificial intelligence and the sensorium of sight. Technol Rev 81: 2-23.

6. Marr D (1982) Vision: A computational investigation into the human representation and processing of information freemen. New York, 51-79.

7. Hubel DH, Wiesel TN (1959) Receptive fields of single neurons in the cat's striate cortex. J Physiol 148: 574-591.

8. Hubel DH, Wiesel TN (1968) Receptive fields and functional architecture of monkey striate cortex. J Physiol 195: 215-243.

9. Ruedi PF, Heim P, Kaess F, et al. (2003) A 128 × 128 pixel 120-db dynamic-range vision-sensor chip for image contrast and orientation extractdion. IEEE Journal of Solid-State Circuits 38: 2325-2333.

10. Choi TYW, Merolla PA, Arthur JV, et al. (2005) Neuromorphic implementation of orientation hypercolumns. IEEE Transactions on Circuits and Systems-I 52: 1049-1060.

11. Rasche C (2007) Neuromorphic excitable maps for visual processing. IEEE Trans Neural Netw 18: 520-529.

12. Serrano-Gotarredona R, Serrano-Gotarredona T, Acosta-Jinenez A, et al. (2008) On real-time aer 2d convolutions

hardware for neuromorphic spike-based cortical processing. IEEE Transactions on Neural Networks 19: 1196-1219.

13. Ferrari V, Ferrari L, Jurie F, et al. (2008) Groups of adjacent contour segments for object detection. IEEE Trans Pattern Anal Mach Intell 30: 36-51.

14. Folowosele F, Vogelstein J, Etienne-Cummings R (2011) Towards a cortical prosthesis: Implementing a spike-based hmax model of visual object recognition in silico. IEEE Journal on Emerging and Selected Topics in Circuits and Systems 1: 516-525.

15. Zamarreno-Ramos C, Camunas-Mesa LA, Perez-Carrasco JA, et al. (2011) On spike-timing-dependent-plasticity, memristive devices, and building a self-learning visual cortex. Front Neurosci 5: e26.

16. Layher G, Brosch T, Neumann H (2017) Real-time biologically inspired action recognition from key poses using a neuromorphic architecture. Front Neurorobot 11: 13.

17. Greene E, Hautus MJ (2017) Demonstrating invariant encoding of shapes using a matching judgment protocol. AIMS Neurosci 4: 120-147.

18. Ahissar E, Arieli A (2012) Seeing via miniature eye movements: A dynamic hypothesis for vision. Front Comput Neurosci 6: 89.

19. Rucci M, Victor JD (2015) The unsteady eye: An information-processing stage, not a bug. Trends Neurosci 38: 195-206.

20. Gollisch T, Meister M (2008) Rapid neural coding in the retina with relative spike latencies. Science 319: 1108-1111.

21. Bullock TH (1993) Integrative systems research in the brain: Resurgence and new opportunities. Annu Rev Neurosci 16: 1-15.

22. Hopfield JJ (1995) Pattern recognition computation using action potential timing for stimulus representation. Nature 376: 33-36.

23. Thorpe S, Fize D, Marlot C (1996) Speed of processing in the human visual system. Nature 381: 520-522.

24. Van Rullen R, Thorpe SJ (2001) Is it a bird? Is it a plane? Ultra-rapid visual categorization of natural and artifactual objects. Perception 30: 655-668.

25. Van Rullen R, Thorpe SJ (2002) Surfing a spike wave down the ventral stream. Vision Res 42: 2593-2615.

26. Kirchner H, Thorpe SJ (2006) Ultra-rapid object detection with saccadic eye movements: Visual processing speed revisited. Vision Res 46: 1762-1776.

27. Wei W, Feller MB (2011) Organization and development of direction-selective circuits in the retina. Trends Neurosci 34: 638-645.

28. Taylor WR, Smith RG (2012) The role of starburst amacrine cells in visual signal processing. Vis Neurosci 29: 73-81.

29. Clark DA, Demb JB (2016) Parallel computations in insect and mammalian visual motion processing. Curr Biol 26: 1062-1072.

30. Mauss AS, Vlasits A, Borst A, et al. (2017) Visual circuits for direction selectivity. Annu Rev Neurosci 40: 211-230.

31. Greene E (2007) Retinal encoding of ultrabrief shape recognition cues. PLoS One 2: e871.

32. Greene E (2016) How do we know whether three dots form an equilateral triangle? JSM Brain Sci 1: 1002.

33. Dacey DM (1989) Axon-bearing amacrine cells of the Macaque monkey retina. J Comp Neurol 284: 275-293.

34. Rodieck RW (1998) The primate retina. In: Steklis HD, Erwin J, Neuroscience-comparative primate biology. Alan R Liss, New York, 203-278.

35. Ammermuller J, Weller R (1988) Physiological and morphological characterization of off-center amacrine cells in the turtle retina. J Comp Neurol 273: 137-148.

36. Famiglietti EV (1992) Polyaxonal amacrine cells of rabbit retina: Morphology and stratification of PA1 cells. J Comp Neurol 316: 391-405.

37. Famiglietti EV (1992) Polyaxonal amacrine cells of rabbit retina: Size and distribution of PA1 cells. J Comp Neurol 316: 406-421.

38. Freed MA, Pflug R, Kolb H, et al. (1996) On-Off amacrine cells in cat retina. J Comp Neurol 364: 556-566.

39. Volgi B, Xin D, Amarillo Y, et al. (2001) Morphology and physiology of the polyaxonal amacrine cells in the rabbit retina. J Comp Neurol 440: 109-125.

40. Wright LL, Vaney DI (2004) The type 1 polyaxonal amacrine cells of the rabbit retina: A tracer-coupling study. Vis Neurosci 21: 145-155.

41. Olveczky BP, Baccus SA, Meister M (2003) Segregation of object and background motion in the retina. Nature 423: 401-408.

42. Roska B, Werblin F (2003) Rapid global shifts in natural scenes block spiking in specific ganglion cell types. Nature Neuroscience 6: 600-608.

43. Ackert JM, Wu SH, Lee JC, et al. (2006) Light-induced changes in spike synchronization between coupled on direction selective ganglion cells in the mammalian retina. J Neurosci 26: 4206-4215.

44. Kenyon GT, Theiler J, George JS, et al. (2004) Correlated firing improves stimulus discrimination in a retinal model. Neural Comput 16: 2261-2291.

45. Miller JA, Denning KS, George JS, et al. (2006) A high frequency resonance in the responses of retinal ganglion cells to rapidly modulated stimuli: A computer model. Vis Neurosci 23: 779-794.

46. Fukushima K (1980) Neocognitron: A self organizing neural network model for a mechanism of pattern recognition unaffected by shift in position. Biol Cybern 36: 193-202.

47. Rolls ET (1992) Neurophysiological mechanisms underlying face processing within and beyond the temporal cortical visual areas. Philos Trans R Soc Lond B Biol Sci 335: 11-20.

48. Wallis G, Rolls ET (1997) Invariant face and object recognition in the visual system. Prog Neurobiol 51: 167-194.

49. Rodriguez-Sanchez AJ, Tsotsos JK (2012) The roles of endstopped and curvature tuned computations in a hierarchical representation of 2d shape. PLoS One 7: e42058.

50. Riesenhuber M, Poggio T (2000) Models of object recognition. Nat Neurosci 3: 1199-1204.

51. Pasupathy A, Connor CE (2001) Shape representation in area V4: Position-specific tuning for boundary conformation. J Neurophysiol 86: 2505-2519.

52. Suzuki N, Hashimoto N, Kashimori Y, et al. (2004) A neural model of predictive recognition in form pathway of visual cortex. Biosystems 76: 33-42.

53. Pinto N, Cox DD, DeCarlo JJ (2008) Why is real-world visual object recognition hard? PLoS Comput Biol 4: e27.